\documentclass[11pt,english]{article}
\usepackage{lmodern}
\usepackage{color}

\usepackage[T1]{fontenc}
\usepackage[utf8x]{inputenc}
\usepackage{amsthm}
\usepackage{amsmath}
\usepackage{amssymb}
\usepackage{esint}
\usepackage{textcomp}
\usepackage{graphicx}
\usepackage{float}
\usepackage{epstopdf}

\makeatletter

\providecommand{\tabularnewline}{\\}

  \theoremstyle{remark}
  \newtheorem{rem}{\protect\remarkname}
  \theoremstyle{plain}
  \newtheorem{prop}{\protect\propositionname}
  \theoremstyle{plain}
  \newtheorem{lem}{\protect\lemmaname}

\@ifundefined{date}{}{\date{}}
\usepackage{a4wide}

\numberwithin{table}{section}
\numberwithin{figure}{section}

\title{Option pricing and hedging with execution costs and market impact\thanks{This research has been conducted with the support of the Research Initiative ``Ex\'ecution optimale et statistiques de la liquidit\'e haute
fr\'equence'' under the aegis of the Europlace Institute of Finance. The
authors want to thank Robert Almgren (Quantitative Brokers), Nicolas Grandchamp des Raux (HSBC France), Charles-Albert Lehalle (CFM), Terry Lyons (Oxford Man), Ramzi Maghrebi (HSBC France), Huyen Pham (Universit\'e Paris-Diderot), Chris Rogers (Cambridge), Mathieu Rosenbaum (UPMC) and Guillaume Royer (Ecole Polytechnique) for the discussions we had on the topic. The authors also want to thank two anonymous referees for their comments on our paper.}}
 \author{Olivier {\sc Gu\'eant} \footnote{Universit\'e Paris-Diderot, UFR de
Math\'ematiques, Laboratoire Jacques-Louis Lions, gueant@ljll.univ-paris-diderot.fr}
 \and Jiang {\sc Pu}  \footnote{Institut Europlace de Finance, jiang.pu.2009@m4x.org }}

\makeatother

\usepackage{babel}
  \providecommand{\lemmaname}{Lemma}
  \providecommand{\propositionname}{Proposition}
  \providecommand{\remarkname}{Remark}

\begin{document}
\maketitle
\noindent
\begin{abstract}
This article considers the pricing and hedging of a
call option when liquidity matters, that is, either for a large nominal
or for an illiquid underlying asset. In practice, as opposed to the classical
assumptions of a price-taking agent in a frictionless market, traders
cannot be perfectly hedged because of execution costs and market impact.
They indeed face a trade-off between hedging errors and costs
that can be solved by using stochastic optimal control. Our modelling framework, which is inspired by the recent literature on optimal execution, makes it possible
to account for both execution costs and the lasting market impact
of trades. Prices are obtained through the indifference pricing approach.
Numerical examples are
provided, along with comparisons to standard methods. \vspace{10mm}

\noindent \textbf{Key words:} Option pricing, Option hedging, Illiquid
markets, Optimal execution, Stochastic optimal control. \vspace{5mm}

\end{abstract}

\section{Introduction}

Classical option pricing theory is based on the hypothesis of a frictionless
market in which agents are price takers: there are no transaction costs and traders have no impact -- neither temporary
nor permanent -- on prices.
These assumptions are not realistic, but the resulting option pricing models
(for instance the Black-Scholes model or the Heston model) are widely used and provide useful results as long as the underlying asset is liquid and the nominal is not too large. However, for options
on illiquid assets or for options with a large nominal compared to the volume commonly traded on the market of the underlying asset, execution
costs and market impact cannot be ignored.

Several improvements to the Black-Scholes model have been made to account for
transaction costs. The basic idea is that high frequency hedging costs are prohibitive due to transaction fees, whereas low frequency hedging leads to large tracking errors. Leland proposed in \cite{tcleland} one of the first models to deal with transaction
costs in the context of option pricing. Other models of frictional markets with either fixed transaction costs or transaction costs proportional to the traded volume include \cite{tcbarlessoner}, \cite{tccvitanic1}, and \cite{tccvitanic2}.

Two other routes have been considered to account for
market imperfections in option pricing models.

The first route is usually referred to as the ``supply curve'' approach.
In this approach, introduced by Çetin, Jarrow and Protter \cite{scetin}
(see also \cite{sbank} and \cite{scetin2,cetin3}), traders are not
price takers, and the price they pay depends on the quantity
they trade. Although appealing, this framework leads to prices identical to those in the Black-Scholes
model. Çetin, Soner and Touzi \cite{cetingamma} consider the same
approach but they restrict the set of admissible strategies (see also
\cite{longstaff}) to obtain positive liquidity costs and prices that eventually
depart from those in the Black-Scholes model. We model execution costs (liquidity
costs) differently because our framework is inspired by the literature
on optimal execution (see \cite{almgreninit,gueant,schied}).

The second route has to do with the impact of $\Delta$-hedging on
the dynamics of the underlying asset,\footnote{This impact was observed in July 2012 through saw-tooth patterns on the prices of five major US stocks (see \cite{lehalle1,lehalle2,lialmgren}).} and the resulting feedback effect
on the price of the option. An important amount of literature exists on this topic, therefore we refer to \cite{fplaten},
\cite{fschon} and \cite{fsircar} for the different modelling approaches. To take account of this effect, we use the same linear form of permanent market impact as in most papers on optimal execution.

In addition to these two routes, a new approach has recently emerged, in relation with the literature on optimal execution.
Rogers and Singh \cite{rogerssingh} and Li
and Almgren \cite{lialmgren} consider approaches inspired by this literature, and similar to ours. In their settings, the authors consider execution costs that are not linear in the volume executed but instead are
convex to account for liquidity effects.

Rogers and
Singh consider an objective function that penalizes both execution
costs and the mean-squared hedging error at maturity. They obtain, in
this close-to-mean-variance framework, a closed form approximation
for the optimal hedging strategy when illiquidity costs are small.

Li and Almgren, motivated by the swings observed in US stock prices
in July 2012, consider a model with both permanent
and temporary impacts, whereas Rogers and Singh do not examine permanent market impact in \cite{rogerssingh}. They use a mean-variance optimization criterion where the hedging error is the main variable. They consider the case of quadratic execution costs and use a constant-$\Gamma$ approximation in order to obtain a closed form expression for the hedging strategy.

Instead of focusing on a special case that leads to closed form expressions, our goal is to consider a general model. We use a general form for the execution costs and we examine the influence of permanent market impact. The optimization criterion we consider is an expected utility applied to final wealth. Therefore, we characterize the optimal strategy with a partial differential equation (PDE), and we rely on numerical methods to approximate the solution of the PDE and the optimal hedging strategy. Another difference is that we account for interest rate. Furthermore, neither \cite{lialmgren} nor \cite{rogerssingh} distinguish physical delivery from cash settlement. In this paper, we show how hedging strategies and option prices are impacted by the type of settlement.

In terms of (partial) hedging strategies, both \cite{lialmgren} and \cite{rogerssingh} obtain optimal strategies that are mean reverting around the classical $\Delta$. This is not the case in our expected utility framework: our optimal strategy does not oscillate around a solution without execution costs and market impact. The optimal strategies in our model are smoother than classical $\Delta$-hedging strategies because the trader seeks to avoid round trips on the stock, which entail execution costs upon purchase and sale of shares. In the case of a physical settlement, smooth strategies are also linked to the fact that the trader is averse both to price risk and to the binary risk of having to deliver versus not delivering. By comparing our strategies with classical $\Delta$-hedging strategies for different frequencies of rebalancing, we show that our approach makes it possible to reach very low levels of variance while mitigating execution costs. In particular, we observe that for the frequency of rebalancing that leads to the same level of execution costs, the variance of the PnL associated with our strategy is lower than the one obtained with a classical $\Delta$-hedging strategy.

In addition to optimal hedging strategies, our expected CARA utility framework provides prices by using the indifference pricing approach. We compute the amount a client needs to pay to compensate, in utility terms, the payoff of the option when the trader uses the optimal hedging strategy. We find that the price of a call is higher in the presence of execution costs than in the classical model, and that this price is an increasing function of the illiquidity of the underlying asset and the nominal of the option.

Although we concentrate on the case of a call option throughout the paper, the same approach can be used for other
types of options. In particular, a similar approach is used to price and hedge Accelerated Share Repurchase contracts (see \cite{gueantpuroyer,jaimungal}). These contracts are Asian-type options with Bermudan-style exercise dates and a physical delivery.

The remainder of the text is organized as follows. In Section 2, we
present the basic hypotheses of our model and we introduce the Hamilton-Jacobi-Bellman
equation associated with the problem. In Section 3, we solve the
control problem without permanent market impact, and we show that the price of the option satisfies a nonlinear PDE. In Section 4, we then show
how our solution can be extended to the case where there is a permanent market impact.
In Section 5, we discuss numerical methods to solve the problem. In Section 6, we present the outcomes from several examples, and we compare our model with the Bachelier model.

\section{Setup of the model}

\subsection{Notations}

We consider a filtered probability space $\left(\Omega,\mathbb{F},\left(\mathcal{F}_{t}\right)_{t\geq0},\mathbb{P}\right)$
that corresponds to the available information on the market, namely the
market price of a stock up to the observation time. For $0\le s<t\le T$,
we denote $\mathcal{P}(s,t)$ the set of $\mathbb{R}$-valued progressively
measurable processes on $[s,t]$.

The problem we consider is a bank (or a trader) selling a call option on a stock to a client.\footnote{The reasoning is the same for a put option or if the bank is buying the option. We consider the specific case of a call option
to highlight the difference between physical delivery and cash settlement.
} The call option has a nominal $N$ (in shares), a strike $K$,
and a maturity $T$.

\paragraph{Execution process}

Because of the execution costs, the bank is not able to replicate
the option. However it buys and sells shares progressively to (partially) hedge its risky position. To model the execution process, we first introduce the market volume process $(V_{t})_{t}$, that is assumed to be deterministic, nonnegative, and bounded. The trading is constrained to not go too fast, relative to the market volume, by imposing a maximum participation rate $\rho_m$.

The number of shares in the hedging portfolio is therefore modelled
as\footnote{$q_{0}$ is the number of shares in the portfolio at inception. In illiquid markets, especially for corporate deals including options,
the buyer of the call may provide an initial number of shares (see the discussion in Section 3). We shall consider
below the case where $q_{0}=0$ and the case where $q_{0}$ is set
to the initial Bachelier $\Delta$ thanks to an initial trade with the buyer.
}:
\begin{eqnarray*}
q_{t} & = & q_{0}+\int_{0}^{t}v_{s}ds,
\end{eqnarray*}
where the stochastic process $v$ belongs to the set of admissible strategies $\mathcal{A}$ defined by:
\begin{eqnarray*}
\mathcal{A} & := & \left\lbrace v\in\mathcal{P}(0,T), |v_t| \le \rho_m V_t, \textrm{a.e. in } (0,T)\times\Omega \right\rbrace.
\end{eqnarray*}

\begin{rem}
The market volume process can be used to model overnight risk by assuming $V_{t}=0$ when the market is closed.
\end{rem}

\paragraph{Price process}

The price process of the underlying asset is defined under the historical probability as an Ito process of the form\footnote{There is no reason to consider a risk-neutral probability in our framework, as one cannot replicate the payoff of a call option because of the execution costs.}:
\begin{eqnarray*}
dS_{t} & = & \mu dt + \sigma dW_{t} + kv_t dt,
\end{eqnarray*}

where $k \ge 0$ models permanent market impact and where $\mu$ is typically a view on the future trend of the underlying asset.

We consider a linear form for the permanent market impact to avoid dynamic arbitrage (see the analysis of Gatheral \cite{gatheral}). The more general framework proposed in \cite{gueantperm} could be another possibility but we believe it is more suited to intraday problems.

\begin{rem}
We consider a drifted Bachelier dynamics for the price instead of the classical Black-Scholes framework. This is also the case in Almgren and Li's paper \cite{lialmgren}. The underlying reason for this choice is that we consider a CARA utility function which is, \emph{a priori}, incompatible with a geometric Brownian motion. The prices we obtain are therefore subject to criticisms when the option maturity is long (see \emph{e.g.}, \cite{st} for a comparison between the Bachelier and the Black-Scholes option pricing models).
\end{rem}

\paragraph{Cash account and execution costs}

The cash account of the bank follows a dynamics linked to the hedging
strategy. It is, in particular, impacted by execution
costs. These execution costs are modelled through the introduction
of a function $L\in C(\mathbb{R},\mathbb{R}_{+})$ that verifies the following conditions:
\begin{itemize}
\item $L(0)=0$,
\item $L$ is an even function,
\item $L$ is increasing on $\mathbb{R}_{+}$,
\item $L$ is strictly convex,
\item $L$ is asymptotically super-linear, that is:
\begin{eqnarray*}
\lim_{\rho\to+\infty}\frac{L(\rho)}{\rho} & = & +\infty.
\end{eqnarray*}
\end{itemize}

For any $v\in\mathcal{A}$, the cash account $X$ evolves as:
\[
dX_{t}=rX_t dt -v_{t}S_{t}dt-V_{t}L\left(\frac{v_{t}}{V_{t}}\right)dt,
\]
where $r$ is the risk-free rate.

\begin{rem}
In applications, $L$ is often a power function of the form $L(\rho)=\eta\left|\rho\right|^{1+\phi}$
with $\phi>0$, or a function of the form $L(\rho)=\eta\left|\rho\right|^{1+\phi}+\psi|\rho|$
with $\phi,\psi>0$ where $\psi$ takes into account proportional costs such as the bid-ask spread or a stamp duty. In particular, the initial Almgren-Chriss framework corresponds to $L(\rho)=\eta\rho^2$ ($\phi=1,\psi=0$).
\end{rem}

\paragraph{Payoff of the option}

At time $T$, we consider either a physical settlement or a cash settlement.

Let us consider the case of a physical settlement. If the option is exercised, then
the bank receives $KN$ and needs to deliver $N$ shares. Because the hedging portfolio contains $q_T$ shares at time $T$, the bank has to buy $N-q_T$ shares to be able to deliver. Thus, if the option is exercised, then the payoff of the bank is:
\begin{eqnarray*}
X_{T}+KN-(N-q_{T})S_{T}-\mathcal{L}(q_{T},N) & = & X_{T}+q_{T}S_{T}+N(K-S_{T})-\mathcal{L}(q_{T},N),
\end{eqnarray*}
where $\mathcal{L}(q,q')$ models the additional cost over the Mark to Market (MtM) price to go from a
portfolio with $q$ shares to a portfolio with $q'$ shares.

In the case where the option is not exercised, the payoff is
\[
X_{T}+q_{T}S_{T}-\mathcal{L}(q_{T},0),
\]
because the trader needs to liquidate the portfolio. The term $\mathcal{L}(q_{T},0)$ is the discount incurred to liquidate the remaining shares.

If we assume that the option is exercised if and only if the stock price is above $K$,\footnote{The threshold might be less
than the strike $K$ in an illiquid market.} then the total payoff in the case of physical settlement is:

$$X_{T}+q_{T}S_{T}- N (S_{T}-K)_+-1_{S_{T}\ge K}\mathcal{L}(q_{T},N)-1_{S_{T}<K}\mathcal{L}(q_{T},0).$$

In the case of a cash settlement, the only difference is when the option is exercised. In that case, the bank pays $N(S_T - K)$ and liquidates its portfolio (with usually a lot of shares). The liquidation leads to the following payoff for the bank:

$$X_{T}+q_{T}S_{T} - N(S_{T}-K)_+ -\mathcal{L}(q_{T},0).$$

In general, for both the cash and the physical settlements, the payoff is therefore of the form

$$X_{T}+q_{T}S_{T} - \Pi(q_T,S_T),$$ where $\Pi(q,S) \ge N (S-K)_+.$

\paragraph{Optimization}

The stochastic optimal control problem we consider is:
\[
\sup_{v\in\mathcal{A}}\mathbb{E}\left[-\exp\left(-\gamma\left(X_{T}+q_{T}S_{T}-\Pi(q_T,S_T)\right)\right)\right],
\]
where $\gamma$ is the absolute risk aversion parameter of the bank.

\begin{rem}
\label{l}
The penalty function $\mathcal{L}$ needs to be specified. When there is no permanent market impact, a natural choice is $\mathcal{L}(q,q') = \ell(|q-q'|)$ where $\ell$ is an increasing and convex function.\footnote{When permanent market impact is taken into account, $\mathcal{L}$ must have a specific form to avoid dynamic arbitrages (see Section 4).} One candidate for $\ell$ is the risk-liquidity premium of a block trade as in \cite{gueant}. Another candidate is the risk-liquidity premium associated with liquidation at some constant participation rate $\rho$ (for instance $\rho_m$):
$$\ell(q) = \int_T^{T'} L(\rho) V_t dt + \frac{\gamma}2 \sigma^2 \int_T^{T'} q_t^2 dt =  \frac{L(\rho)}{\rho} |q| + \frac{\gamma}2 \sigma^2 \int_T^{T'} \left(|q|- \rho\int_T^t V_s ds\right)^2 dt,$$ where $T'$ is the first time such that $\int_T^{T'} \rho V_t dt = |q|$.\\
\end{rem}

\subsection{The value function and the HJB equation}

To solve the above stochastic optimal control problem, we define the value function $u$ by:
\begin{eqnarray*}
u(t,x,q,S) & = & \sup_{v\in\mathcal{A}_{t}}\mathbb{E}\left[-\exp\left(-\gamma\left(X_{T}^{t,x,v}+q_{T}^{t,q,v}S_{T}^{t,S,v}- \Pi(q_{T}^{t,q,v},S_{T}^{t,S})\right)\right)\right],
\end{eqnarray*}

where:
\begin{eqnarray*}
\mathcal{A}_{t} & := & \left\lbrace v\in\mathcal{P}(t,T), |v_s| \le \rho_m V_s, \textrm{a.e. in } (t,T)\times\Omega\right\rbrace ,
\end{eqnarray*}
and where:
\begin{eqnarray*}
X_{t'}^{t,x,v} & = & x+\int_{t}^{t'}\left(rX^{t,x,v}_s-v_{s}S_{s}^{t,S,v}-V_{s}L\left(\frac{v_{s}}{V_{s}}\right)\right)ds\\
q_{t'}^{t,q,v} & = & q+\int_{t}^{t'}v_{s}ds\\
S_{t'}^{t,S,v} & = & S+\mu (t'-t)+\int_{t}^{t'}\sigma dW_{s} +k(q^{t,q,v}_{t'}-q).
\end{eqnarray*}
The Hamilton-Jacobi-Bellman (HJB) equation associated with this problem
is the following:
\begin{eqnarray*}
-\partial_{t}u- \mu \partial_S u-\dfrac{1}{2}\sigma^{2}\partial_{SS}^{2}u-\sup_{|v|\le \rho_m V_t}\left\{ v\partial_{q}u+\left(rx-vS-L\left(\dfrac{v}{V_{t}}\right)V_{t}\right)\partial_{x}u+kv\partial_Su\right\}  & = & 0,
\end{eqnarray*}

with the terminal condition:
\begin{eqnarray*}
u(T,x,q,S) & = & -\exp\left(-\gamma\left(x+qS-\Pi(q,S)\right)\right).
\end{eqnarray*}

\begin{rem}
In the case of a physical settlement, this terminal condition is not continuous.
\end{rem}

\section{Characterization of the solution}

We first consider the case without permanent market impact $(k=0)$. In that case, we consider a function $\mathcal{L}$ of the form $\mathcal{L}(q,q')=\ell(|q'-q|)$,
where $\ell$ is a convex and even function that increases on $\mathbb{R}_{+}$ (as exemplified in Remark \ref{l}).

The following lemma states that we can factor out the compounded MtM value of the current portfolio (we omit the superscripts to improve readability):

\begin{lem}
\label{xqS}
$$X_{T}+q_{T} S_{T} = e^{r(T-t)}(x+qS)$$$$ +  e^{r(T-t)} \left(\int_t^T e^{-r(s-t)} q_s(\mu-rS_s)ds + \int_t^T e^{-r(s-t)} q_s \sigma dW_s - \int_t^T e^{-r(s-t)} V_{s}L\left(\dfrac{v_{s}}{V_{s}}\right) ds  \right).$$
\end{lem}

This lemma shows that the value function  $u(t,x,q,S)$  is of the form:

\begin{eqnarray*}
u(t,x,q,S) & = & -\exp\left(-\gamma e^{r(T-t)}(x+qS)\right)\inf_{v\in\mathcal{A}_{t}}J_{t}(q,S,v).
\end{eqnarray*}
where
\begin{eqnarray*}
J_{t}:\mathbb{R}\times\mathbb{R}\times\mathcal{A}_{t} & \to & \mathbb{R}\\
(q,S,v) & \mapsto & J_{t}(q,S,v)
\end{eqnarray*}
is defined as
\begin{eqnarray*}
J_{t}(q,S,v) & = & \mathbb{E}\left[\exp\left(-\gamma\left(e^{r(T-t)}\left(\int_t^T e^{-r(s-t)} q_s(\mu-rS_s)ds + \int_t^T e^{-r(s-t)} q_s \sigma dW_s\right.\right.\right.\right.\\
                    & & \left.\left.\left.\left.- \int_t^T e^{-r(s-t)} V_{s}L\left(\dfrac{v_{s}}{V_{s}}\right) ds  \right)-\Pi(q_T,S_T))\right)\right)\right].
\end{eqnarray*}
We also define
\begin{eqnarray*}
\theta(t,q,S) & = & \inf_{v\in\mathcal{A}_{t}}\frac{e^{-r(T-t)}}{\gamma}\log(J_{t}(q,S,v)).
\end{eqnarray*}

The following proposition states that $\theta$ is well defined, and gives a lower bound for $\theta$:
\begin{prop}
\label{bounds} $\forall(t,q,S)\in[0,T]\times\mathbb{R}\times\mathbb{R}$, $\theta(t,q,S)$ is finite.\\
Moreover, if $\mu=r=0$ then:
\begin{eqnarray*}
\theta(t,q,S) & \ge & N\mathbb{E}\left[(S_{T}-K)_+\right].
\end{eqnarray*}
\end{prop}

The function $\theta$ has a natural interpretation. Let us consider a call option deal between the bank and a client, where at time $0$:
\begin{itemize}
\item the bank writes the call option with either a physical or cash settlement and the client pays a price $P$, and
\item the client gives $q_0$ shares to the bank and receives $q_0 S_0$ in cash from the bank.
\end{itemize}

In utility terms, the bank gives the following value to this deal $$u(0,X_0 - q_0 S_0+P,q_0,S_0) = -\exp\left(-\gamma e^{rT}\left(X_0+P-\theta(0,q_0,S_0)\right)\right).$$

As a consequence, if $P = \theta(0,q_0,S_0)$, and if we assume that the cash is invested at rate $r$, then the bank is indifferent to making the deal or not making the deal. Therefore, $\theta(0,q_0,S_0)$ is the indifference price of the call option deal.

This definition of a price for the call option depends on $q_0$: the initial
number of stocks in the portfolio. This echoes the fact that, in practice, building the initial $\Delta$ position (as computed in a classical model)  is
usually costly for options with a large nominal.

This interpretation of $\theta$ also enables us to see the inequality of Proposition \ref{bounds}
in a different manner. When $\mu=r=0$, then the price in our setting is always
greater than the price when there is no execution cost (Bachelier model).%
\footnote{The price of a call (with a unitary nominal and when $r=0$) in the Bachelier
model is given by:
\begin{eqnarray*}
\mathbb{E}\left[\left(S_{T}^{t,S}-K\right)_+\right] & = & (S-K)\Phi\left(\dfrac{S-K}{\sigma\sqrt{T-t}}\right)+\sigma\sqrt{T-t}\varphi\left(\frac{S-K}{\sigma\sqrt{T-t}}\right),
\end{eqnarray*}
where $\varphi$ and $\Phi$ are respectively the probability density
function and the cumulative distribution function of a standard normal
variable.%
}

Our focus now is on the function $\theta$ because it is the price of the call option. Our first result on $\theta$ states that it is a convex function of $q$:

\begin{prop}
\label{convex}
For $(t,S)\in[0,T]\times\mathbb{R},\qquad q\in\mathbb{R}\mapsto\theta(t,q,S)$
is a convex function.\end{prop}

\begin{rem}
We cannot expect the same result for $S$ because the final payoff is not continuous in $S$ in general (see the physical delivery case).
\end{rem}

The main property for $\theta$ is the following PDE characterization:

\begin{prop}
\label{withoutperm} Let us introduce $$H(p) = \sup_{|\rho|\le \rho_m} p\rho - L(\rho).$$
$\theta$ is a viscosity solution of the following equation:
\begin{eqnarray*}
-\partial_{t}\theta + r\theta + (\mu - rS) q -\mu \partial_S \theta -\dfrac{1}{2}\sigma^{2}\partial_{SS}^{2}\theta-\dfrac{1}{2}\gamma\sigma^{2}e^{r(T-t)}(\partial_{S}\theta-q)^{2}+V_{t}H(\partial_{q}\theta)=0,
\end{eqnarray*}
with the terminal condition $\theta(T,q,S)=\Pi(q,S)$
in the classical sense. \end{prop}

The PDE satisfied by $\theta$ is a nonlinear equation and, in particular, the price of the call
option is not proportional to the nominal. To go from a nominal equal
to $N$ to a nominal equal to $1$, we introduce the function
$\tilde{\theta}$ defined by:
\begin{eqnarray*}
\tilde{\theta}(t,\tilde{q},S) & = & \frac{1}{N}\theta(t,N\tilde{q},S).
\end{eqnarray*}
Then, $\tilde{\theta}$ satisfies the following
equation in the viscosity sense:
\begin{eqnarray*}
-\partial_{t}\tilde{\theta} + r \tilde{\theta} +  (\mu - rS) \tilde q  - \mu \partial_S \tilde{\theta}-\dfrac{1}{2}\sigma^{2}\partial_{SS}^{2}\tilde{\theta}-\dfrac{1}{2}\gamma N\sigma^{2}e^{r(T-t)}(\partial_{S}\tilde{\theta}-\tilde{q})^{2}+\frac{V_{t}}{N}H(\partial_{\tilde{q}}\tilde{\theta}) & = & 0,
\end{eqnarray*}
with the terminal condition
\begin{eqnarray*}
\tilde{\theta}(T,\tilde{q},S) & = & \frac{1}{N}\Pi(N\tilde{q},S).
\end{eqnarray*}
In other words, we need to rescale the risk aversion parameter $\gamma$, the
market volume process $(V_t)_t$, and the liquidation penalty function $\mathcal{L}$ in order to go
from a call of nominal $N$ to a call of nominal $1$.

Each term in the PDE

$$\partial_{t}{\theta} = \underbrace{r {\theta}}_{\textrm{(I)}} +  \underbrace{(\mu - rS)  q}_{\textrm{(II)}}  \underbrace{- \mu \partial_S {\theta}-\dfrac{1}{2}\sigma^{2}\partial_{SS}^{2}{\theta}}_{\textrm{(III)}}-\underbrace{\dfrac{1}{2}\gamma \sigma^{2}e^{r(T-t)}(\partial_{S}{\theta}-{q})^{2}}_{\textrm{(IV)}}+\underbrace{V_{t}H(\partial_{{q}}{\theta})}_{\textrm{(V)}}$$

has a specific interpretation:

\begin{itemize}
\item The term (I) is the classical term linked to discounting at risk-free rate $r$.
\item The term (II) corresponds to the premium linked to holding shares instead of cash. If indeed one holds $q$ shares, on average the MtM wealth is increased by $\mu q$ per unit of time, whereas the amount of cash equivalent to $q$ shares (that is $qS$) increases the MtM wealth by $rqS$ per unit of time.
\item  The term (III) is linked to the dynamics of the stock price.
\item The interdependence between the number of shares $q$ in the hedging
portfolio and the dynamics of the price occurs through (IV), and more precisely through the term
$(\partial_{S}\theta-q)^{2}$. Although there is no $\Delta$ in this
model because the market is incomplete, this term measures the difference
between the first derivative of the option price with respect to the
price of the underlying asset and the number of shares in the hedging portfolio: it looks therefore
like the measure of a mis-hedge.
\item And, (V) is the classical term in the literature on optimal execution. It models the execution costs and the participation limit $\rho_m$. In particular, the optimal participation rate at time $t$ is $\rho^*(t,q_t,S_t) = \frac{v^*(t,q_t,S_t)}{V_t} = H'(\partial_q\theta(t,q_t,S_t))$.
\end{itemize}

\begin{rem}
If we replace $\mu$ with $rS$ and $\sigma$ with $\sigma S$, then the terms (I), (II) and (III) are exactly the same as those in the Black-Scholes PDE.
\end{rem}

Furthermore, the partial differential equation satisfied by $\theta$
is (surprisingly) not derived from a control problem because
\begin{eqnarray*}
(p_{q},p_{S}) & \mapsto & -\dfrac{1}{2}\gamma\sigma^{2}e^{r(T-t)}(p_{S}-q)^{2}+V_{t}H(p_{q})
\end{eqnarray*}
is neither convex, nor concave.

In fact, it derives from a zero-sum game (see the appendix of \cite{bardi})
in which the first player controls $q$ through
\begin{eqnarray*}
dq_{t} & = & v_{t}dt,
\end{eqnarray*}
and player 2 controls the drift of the price
\begin{eqnarray*}
dS_{t} & = & (\mu + \alpha_t)dt+\sigma dW_{t}.
\end{eqnarray*}
The payoff of the zero-sum game associated with the above equation is:
$$
\mathbb{E}\Bigg[\int_{0}^{T}e^{r(T-t)}\Bigg(L\left(\frac{v_{t}}{V_{t}}\right)V_{t}-\frac{e^{-r(T-t)}}{2\gamma\sigma^{2}}(\alpha_{t}+\gamma e^{r(T-t)}\sigma^{2}q_{t})^{2}$$$$-\frac{1}{2}\gamma e^{r(T-t)}\sigma^{2}q_{t}^{2} + (rS_t-\mu)q_t\Bigg)dt + \Pi(q_T,S_T)\Bigg],
$$
where player 1 minimizes and player 2 maximizes.

\section{The problem with permanent market impact}

We now turn to the case where there is a permanent market impact. To stay in the framework of Gatheral's paper on permanent market impact without dynamic arbitrage \cite{gatheral}, we consider in this section that $\mu=r = 0$, and we use the linear form of permanent market impact introduced in Section 2. We will show that, up to a change of variables, the problem is -- from a mathematical point of view -- the same as in the absence of permanent market impact.

To avoid dynamic arbitrage we have to specify $\mathcal{L}$. At time
$T$, if one wants to go from a portfolio with $q$ shares
to a portfolio with $q'$ shares, one must pay the liquidity costs related to the volume transacted. This is modelled by $\ell(q'-q)$, as in
the previous case without permanent market impact. However, we must also
take permanent market impact into account.
The amount paid to go from a portfolio with $q$ shares
at time $T$ to a portfolio with $q'$ at time $T'$ is (on average
and ignoring temporary market impact):
\begin{eqnarray*}
\mathbb{E}\left[\left.\int_{T}^{T'}dq_{t}S_{t}\right|S_{T}\right] & = & (q'-q)S_{T}+\int_{T}^{T'}k(q'-q_{t})dq_{t}\\
 & = & (q'-q)S_{T}+\frac 1 2 k (q'-q)^2.\\
\end{eqnarray*}

Hence, we define $\mathcal{L}$ by:
\begin{eqnarray*}
\mathcal{L}(q,q') & = & \ell(|q'-q|)+\frac 12 k (q'-q)^2.
\end{eqnarray*}

Let us now come to the change of variables. In the previous section, we showed that $u(t,x,q,S)$,
in the absence of permanent market impact, can be written as:
\begin{eqnarray*}
u(t,x,q,S) & = & -\exp\left(-\gamma\left(x+qS-\theta(t,q,S)\right)\right).
\end{eqnarray*}
Using the same method as in Section 3, we show that, with permanent market
impact, $u$ can be written as:
\begin{eqnarray*}
u(t,x,q,S) & = & -\exp\left(-\gamma\left(x+qS-\frac 12 k q^2 + \frac 12 k q_0^2 -\theta(t,q,S-k(q-q_0))\right)\right).
\end{eqnarray*}

In other words, we introduce the function:
\begin{eqnarray*}
\theta(t,q,\tilde{S}) & = & x+q(\tilde{S}+k(q-q_0))-\frac 12 k q^2 + \frac 12 k q_0^2+\frac{1}{\gamma}\log\left(-u(t,x,q,\tilde{S}+k(q-q_0))\right).
\end{eqnarray*}

As in the previous case, $\theta(0,q_{0},S_{0})$ is the price of
the call at time $0$ when the deal starts with an exchange of $q_{0}$ shares against $q_0S_0$ in cash.

\begin{rem}
The new variable $\tilde{S}_{t}=S_{t}-k(q_t-q_0)$ is the price from which we remove
the influence of the permanent market impact.
\end{rem}
Now, using the same techniques as in Section 3, we prove the following
proposition:
\begin{prop}
\label{withperm} Let us assume that $\mu=r=0$. Then, $\theta$ is a viscosity solution of the following equation:
\begin{eqnarray*}
-\partial_{t}\theta-\dfrac{1}{2}\sigma^{2}\partial_{\tilde{S}\tilde{S}}^{2}\theta-\dfrac{1}{2}\gamma\sigma^{2}(\partial_{\tilde{S}}\theta-q)^{2}+V_{t}H(\partial_{q}\theta) & = & 0,
\end{eqnarray*}
with the terminal condition
\begin{eqnarray*}
\theta(T,q,\tilde{S}) & = & \Pi(q,\tilde{S}+k(q-q_0)) - \frac 12 kq^2 + \frac 12 kq_0^2\\
\end{eqnarray*}
\end{prop}

The introduction of a permanent market impact only changes the terminal condition of the PDE. In the case of a cash settlement, the terminal condition is:
$$\theta(T,q,\tilde{S}) = N(\tilde{S}+k(q-q_0)-K)_+ + \ell(q) + \frac 12 k q_0^2.$$
In the case of a physical settlement, the terminal condition is:
$$\theta(T,q,\tilde{S}) = N(\tilde{S}+k(q-q_0)-K)_+$$$$ + 1_{\tilde{S}+k(q-q_0) \ge K} \left(\ell(N-q) + \frac 12 k N(N-2q)\right)+ 1_{\tilde{S}+k(q-q_0)< K}\ell(q) +\frac 12 k q_0^2.$$

\section{Numerical methods}

We now present two numerical methods to approximate the solution of our hedging and pricing problem.

\subsection{Numerical method for the PDE}

Proposition \ref{withoutperm} and Proposition \ref{withperm} show
that the hedging and pricing problem boils down to solving a partial differential equation
in dimension 3. Factoring out the nominal of the call, the PDE is:

\begin{eqnarray*}
\partial_{t}\tilde{\theta} & = & \underbrace{r \tilde{\theta} +  (\mu - rS) \tilde q  - \mu \partial_S \tilde{\theta} -\dfrac{1}{2}\sigma^{2}\partial_{SS}^{2}\tilde{\theta}}_{\textrm{(A)}} - \underbrace{\dfrac{1}{2}\gamma N\sigma^{2}e^{r(T-t)}(\partial_{S}\tilde{\theta}-\tilde{q})^{2}}_{\textrm{(B)}}+\underbrace{\frac{V_{t}}{N}H(\partial_{\tilde{q}}\tilde{\theta})}_{\textrm{(C)}},
\end{eqnarray*}

with a final condition that depends on the nature of the settlement and on whether or not we consider permanent market impact (if the permanent market impact is taken into account, then we consider $\mu = r = 0$).

To approximate a solution of this PDE with the terminal condition corresponding to our problem, we first split the equation into three parts to consider a numerical scheme based on operator splitting. For (A), we consider an implicit finite difference scheme. We always start with this step that smoothes the terminal condition because there is a singularity at time $T$. For (B), we use a monotonic explicit scheme \emph{à la} Godunov, except at the boundaries (see below). For (C), we use a semi-Lagrangian method because it provides the optimal control directly (see \cite{ferretti} for more details on the classical numerical methods for HJB equations). Regarding the boundary conditions, we can use a grid for $q$ that is sufficiently large to search for an optimum inside the domain. However, this is not the case as far as the finite differences in $S$ are concerned (this is related to the fact that $\theta$ describes a zero-sum game). Therefore, we need to specify boundary conditions for the minimum and maximum values of $S$ ($S_{\textrm{min}}$ is assumed to be far below the strike $K$, and $S_{\textrm{max}}$ is assumed to be far greater than $K$). Because $\theta$ is the price of the call option, a natural condition is $\partial_{SS} \theta = 0$ at $S_{\textrm{min}}$ and $S_{\textrm{max}}$. However, this condition leads to a globally non-monotone scheme (see the seminal paper by Crandall and Lions on monotone schemes \cite{mono}). In practice, this scheme provides good results (see below). However, because it requires setting boundary conditions that are not exactly in line with the underlying financial problem, we also develop an alternative method.

\subsection{Tree-based approach}

The above numerical method is based on a finite difference scheme and therefore requires artificially setting the boundary conditions. A way to avoid setting the boundary conditions is to use a tree-based approach. The underlying idea is to discretize the problem and to use the same change of variables as in the continuous model to obtain a way to approximate $\theta$.

We consider the subdivision $t_0 = 0, \ldots, t_j = j\Delta t, \ldots, t_J = J\Delta t = T$. We also consider the sequences $(X_j)_j$, $(q_j)_j$, $(S_j)_j$ defined by the following equations:

\begin{itemize}
\item $S_{j+1} = S_j + \mu \Delta t  + \sigma \sqrt{\Delta t} \epsilon_{j+1},$ where the $\epsilon_j$s are i.i.d. with $\mathbb{E}[\epsilon_j] =0$ and $\mathbb{V}[\epsilon_j] =1$,

\item $q_{j+1} = q_j + v_j \Delta t,$

\item $X_{j+1} = e^{r\Delta t}X_j - v_jS_j \Delta t - L\left(\frac{v_j}{V_{j+1}}\right) V_{j+1} \Delta t.$

\end{itemize}

Our goal is to maximize over $\left\lbrace (v_j)_{0 \le j < J}, |v_j| \le \rho_m V_{j+1} \right\rbrace$, the following expression:

$$\mathbb{E}\left[-\exp\left(-\gamma\left(X_J+q_JS_J - \Pi(q_J,S_J)\right)\right)\right].$$

For that purpose we introduce the value functions:

$$u_j(x,q,S) = \mathbb{E}\left[-\exp\left(-\gamma\left(X_J+q_JS_J - \Pi(q_J,S_J)\right)\right)|X_j = x, q_j = q, S_j= S \right].$$

The Bellman equation associated with the problem is:

$$\forall j \in \lbrace 0, \ldots, J-1\rbrace,\qquad  u_j(x,q,S) =$$$$ \sup_{|v|\le \rho_m V_{j+1}}\mathbb{E}\left[u_{j+1}( e^{r\Delta t}x - vS \Delta t - L\left(\frac{v}{V_{j+1}}\right) V_{j+1} \Delta t, q + v\Delta t, S + \mu \Delta t + \sigma \sqrt{\Delta t} \epsilon_{j+1})\right],$$ and
$$u_J(x,q,S) = -\exp\left(-\gamma\left(x+qS - \Pi(q,S)\right)\right).$$

If we write $u_j(x,q,S) = -\exp\left(-\gamma e^{r(J-j)\Delta t }\left(x+qS - \theta_j(q,S)\right)\right)$, then the Bellman equation becomes:

$$\theta_j(q,S) = \frac{e^{-r(J-j)\Delta t }}{\gamma} \inf_{|v|\le \rho_m V_{j+1}} \log \mathbb{E}\Bigg[\exp\Bigg(\gamma e^{r(J-(j+1))\Delta t}\Bigg( qS(e^{r\Delta t} - 1) + L\left(\frac v{V_{j+1}}\right)V_{j+1}\Delta t$$$$ - (q+v\Delta t)(\mu \Delta t + \sigma \sqrt{\Delta t} \epsilon) + \theta_{j+1}(q+v\Delta t, S+\mu\Delta t+ \sigma \sqrt{\Delta t} \epsilon )\Bigg)\Bigg)\Bigg], \quad 0\le j < J,$$
and
$$\theta_J(q,S) = \Pi(q,S).$$

We now consider a trinomial tree that corresponds to $$\epsilon = \begin{cases} \alpha &\mbox{with probability } \frac{1}{2\alpha^2} \\
0 & \mbox{with probability } 1-\frac{1}{\alpha^2}\\
-\alpha & \mbox{with probability } \frac{1}{2\alpha^2}, \end{cases}$$ where $\alpha >1$.\footnote{In examples, we consider $\alpha = \sqrt{2}$.}\\

Each node of the tree represents a pair $(j,S_j)$ where $S_j \in \lbrace S_0 +\mu j \Delta t + \sigma \sqrt{\Delta t} p\alpha,  -j\le p \le j\rbrace$,  and the tree is naturally recombining because the drift is constant and the noise symmetrical. At a given node $(j,S_j)$, we compute the value of $\theta_j(q,S_j)$ for $q$ on a specified grid (the natural boundaries when $\mu= r= 0 $ are $q_{\textrm{min}} = 0$ and $ q_{\textrm{max}} = N$ for a call option) by using the above recursive equations. In particular, if the market volume is assumed to be constant (equal to $V$), then the step $\Delta q$ of the grid in $q$ should be such that $\rho_m V$ is a multiple of $\Delta q$.

Recursively, by backward induction, we end up at node $(0,S_0)$ with the price of the call for any $q_0$. Also, we get the optimal strategy at each node as a function of $q$.

This method is preferable over the first one because there is no issue with respect to the boundaries in $S$. However, like all tree methods, it ignores the risk of an important price move over a short period of time.

\section{Numerical examples and comparison with the Bachelier model}

\subsection{Examples without permanent market impact}

To exemplify the use of our model and the effectiveness of our numerical methods, we consider the following reference
scenario with no permanent market impact. This reference scenario corresponds to rounded
values for the stock Total SA (the most important component of the CAC 40 Index):
\begin{itemize}
\item $S_{0}=45$ €,
\item $\sigma=0.6$ €$\cdot\text{day}^{-1/2}$ -- it corresponds to an
annual volatility approximately equal to $21\%$,
\item $T=63$ days,
\item $V=4\ 000\ 000$ shares$\cdot$$\text{day}^{-1}$,
\item $N=20\ 000\ 000$ shares,
\item $L(\rho)=\eta|\rho|^{1+\phi}$ with $\eta=0.1$ € $\cdot\mbox{stock }^{-1}\cdot\text{day}^{-1}$, and $\phi=0.75$.
\end{itemize}

For our reference case, we consider $\mu = r= 0$.

Our choice for the risk aversion parameter is $\gamma=2\cdot10^{-7}$ €$^{-1}$.

We consider a call option with strike $K=45$ (at-the-money call option).

Also, we consider (by default) the case where $\rho_m = 500\%$ so that, in practice, there is no constraint on the participation rate. For the terminal cost, we use the form presented in Remark 4 with a participation rate equal to $\rho_m$.

\begin{figure}[H]
\centering \includegraphics[width=0.72\textwidth]{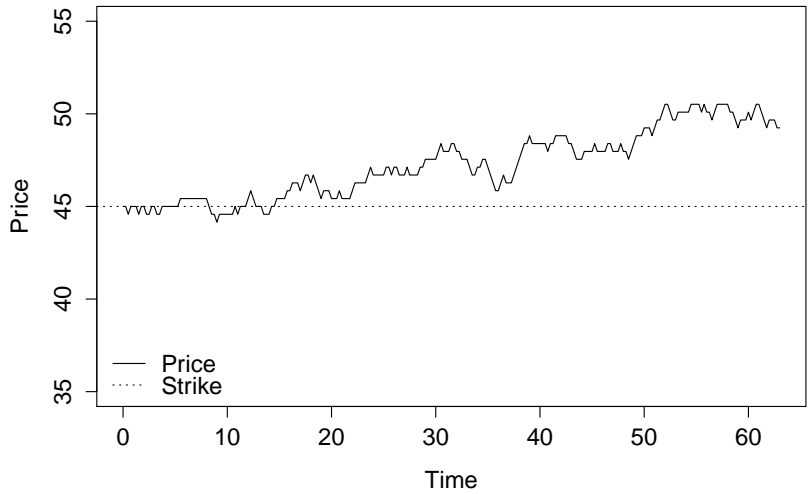} \caption{Trajectory of the stock price.}
\label{price}
\end{figure}

\begin{figure}[H]
\centering \includegraphics[width=0.72\textwidth]{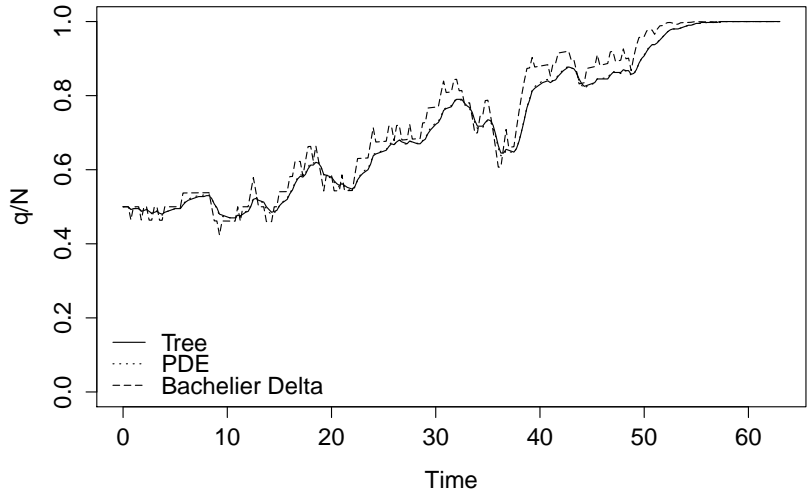} \caption{Results with the tree-based method and the finite difference scheme (PDE).}
\label{methods}
\end{figure}

Figure \ref{methods} presents the outcomes when $q_0 = 0.5 N$ (initial Bachelier $\Delta$) in the case of a physical delivery for the trajectory of the stock price (compatible with the structure of the tree with four levels of nodes per day) represented in Figure \ref{price}. This trajectory corresponds to an exercise of the option at time $T$. Similar results could be obtained in the case of $S_T<K$.

We use the two numerical methods presented above to illustrate the optimal strategy, and we also plot the Bachelier $\Delta$ as a benchmark. Figure \ref{methods} shows that the finite difference scheme and the tree-based method provide almost identical results as far as the optimal strategy is concerned. Moreover, this optimal strategy is different from the hedging strategy in a Bachelier model. As opposed to other papers in the literature, the optimal strategy in our model does not oscillate around the Bachelier $\Delta$ but instead is conservative. Our strategy is smoother because the trader avoids buying too many shares to avoid selling them afterwards, due to the execution costs.

In terms of prices, the results obtained with the PDE method and with the tree-based approach are very close (Table \ref{tabmethods}). However, the difference between the Bachelier price and the price in our approach is significant.\footnote{We divide the prices by $N$ to obtain meaningful figures.}

\begin{table}[H]
\label{tabmethods}
\centering
\begin{tabular}{|c|c|c|c|}
\hline
Model/Method & Bachelier  & Tree-Based approach  & PDE approach \tabularnewline
\hline
Price & 1.900  & 2.060  & 2.067 \tabularnewline
\hline
\end{tabular}
\caption{Prices of the call option for the two numerical methods.}
\end{table}

\subsection{Influence of the parameters}

We now illustrate the main drivers of the difference between our approach and the Bachelier model.

\paragraph{Execution costs}

\begin{figure}[H]
\centering \includegraphics[width=0.68\textwidth]{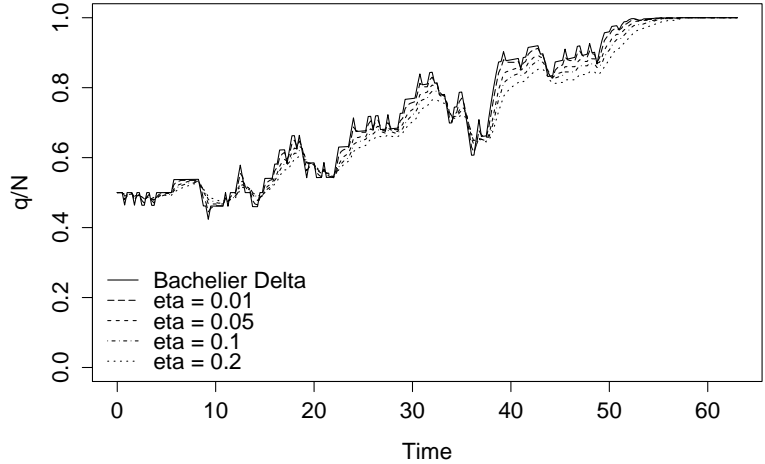} \caption{Optimal strategies for different values of $\eta$ (tree-based approach).}
\label{eta}
\end{figure}

First, because there are execution costs in the model, we illustrate the role of $\eta$. We consider the previous scenario with a physical delivery but with $\eta \in  \lbrace 0.01,0.05, 0.1, 0.2 \rbrace$. The results obtained with the tree-based approach are shown in Figure \ref{eta}.

The effect of execution costs is clear: the higher the execution costs, the smoother the optimal strategy. The trader wants to avoid costly erratic changes in his or her portfolio because of execution costs. Furthermore, the
optimal portfolio gets closer to $0.5N$ when the execution costs increase. This is the same idea as above: because the trader does not know whether
he or she will eventually have to deliver $N$ shares or $0$, he or she
wants to avoid round trips. Therefore, the trader stays closer to $0.5N$ when the liquidity of the underlying asset
decreases.

Table \ref{tabeta} shows that the price of the call increases with $\eta$, as expected.

\begin{table}[H]
\label{tabeta}
\centering
\begin{tabular}{|c|c|c|c|c|c|}
\hline
$\eta$  & 0.2  & 0.1  & 0.05  & 0.01  & 0 (Bachelier) \tabularnewline
\hline
Price of the call  &  2.144 & 2.060  & 2.007  & 1.943  & 1.900 \tabularnewline
\hline
\end{tabular}
\caption{Prices of the call option for different levels of liquidity  (tree-based approach).}
\end{table}

\paragraph{Initial position}

\begin{figure}[H]
\centering \includegraphics[width=0.66\textwidth]{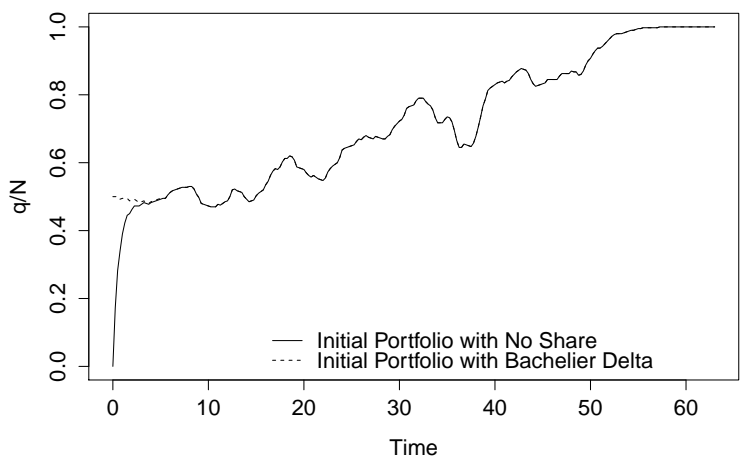} \caption{Optimal portfolio for different values of $q_0$  (tree-based approach).}
\label{init}
\end{figure}

\begin{figure}[H]
\centering \includegraphics[width=0.66\textwidth]{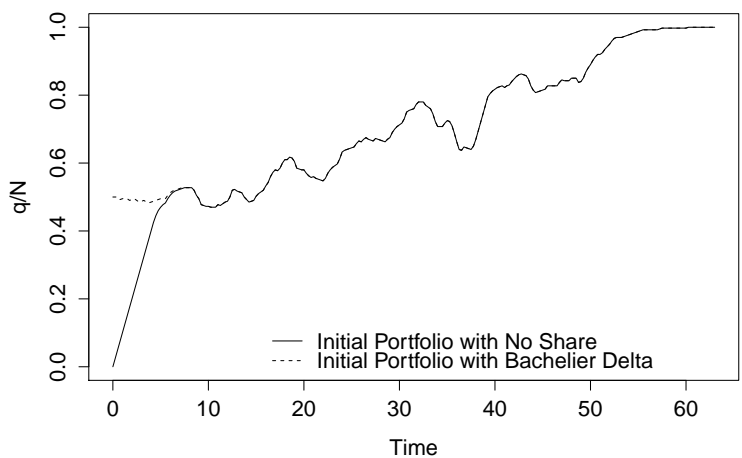} \caption{Optimal portfolio for different values of $q_0$ when $\rho_m = 50\%$  (tree-based approach).}
\label{initc}
\end{figure}

Another parameter linked to liquidity is $q_0$. To understand the role of the initial number of shares, we show in Figure \ref{init} the optimal strategies for $q_0= 0$ and $q_0=0.5N$. To be even more realistic, we add a participation constraint $\rho_m = 50\%$ in Figure \ref{initc}.

The associated prices are given in Table 6.3.

\begin{table}[H]
\label{tabq0}
\centering
\begin{tabular}{|c|c|c|c|c|}
\hline
Values of the parameters & $q_0=0$  & $q_0=0, \rho_m = 50\%$  & $q_0=0.5N$  & $q_0=0.5N, \rho_m = 50\%$ \tabularnewline
\hline
Price of the call  &  2.182 & 2.653  & 2.060  &  2.100\tabularnewline
\hline
\end{tabular}
\caption{Prices of the call option for different values of the initial portfolio and different participation constraints  (tree-based approach).}
\end{table}

Table 6.3 shows that there is a substantial difference between the price of
the call option when $q_{0}=0$ and when $\frac{q_{0}}{N}=0.5$, especially when a participation constraint is imposed. The
rationale for this difference is the cost of building a position consistent
with the risk linked to the option. This is clearly seen in Figures
\ref{init} and \ref{initc}. The two portfolios are
almost the same after a few days. However, the first few days are used by
the trader to buy shares in order to obtain a portfolio close to the
portfolio he or she would have had, had he started with the $\Delta$ in
the Bachelier model.

\paragraph{Price risk and risk aversion}

\begin{figure}[H]
\centering \includegraphics[width=0.68\textwidth]{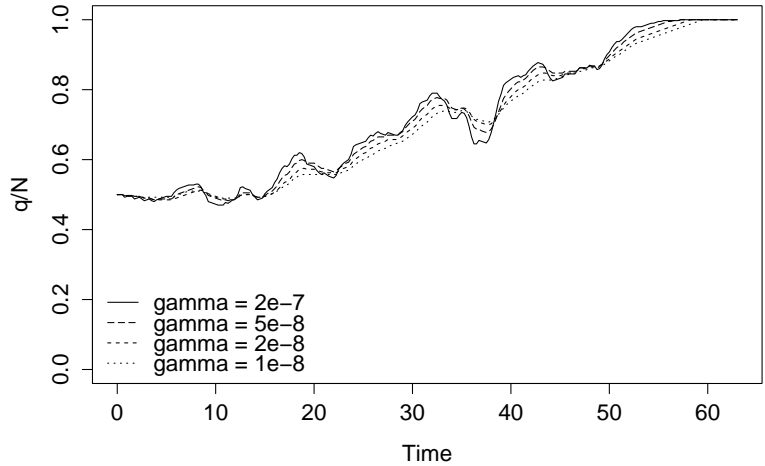} \caption{Optimal portfolio for small values of $\gamma$  (tree-based approach).}
\label{gamma1}
\end{figure}

\begin{figure}[H]
\centering \includegraphics[width=0.68\textwidth]{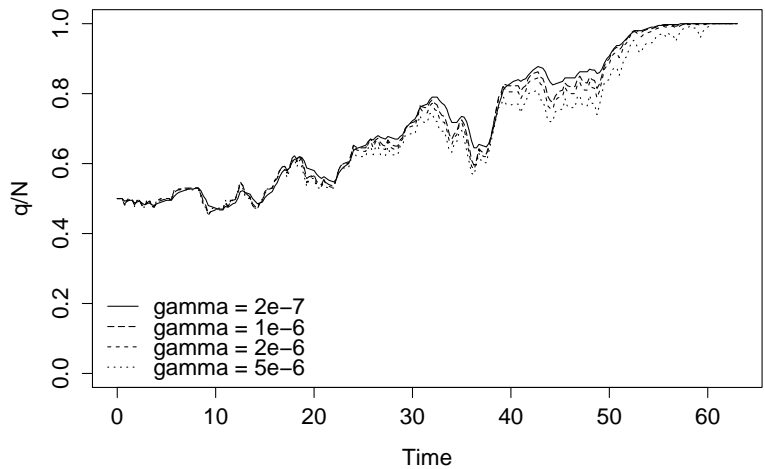} \caption{Optimal portfolio for large values of $\gamma$  (tree-based approach).}
\label{gamma2}
\end{figure}

One of the main parameters when dealing with options is volatility. Here, the influence of the parameter $\sigma$ is clear. The more volatile the stock, the closer to the Bachelier $\Delta$ the hedging strategy is. Also, the price of a call is an increasing function of $\sigma$. What is interesting when it comes to risk is not $\sigma$ but $\gamma$, the risk aversion parameter, because there are two risks of two different natures:

\begin{itemize}
\item The first risk is linked to the optional dimension of the contract: the trader has to deliver either $N$ shares or none. Being averse to this
risk encourages the trader to stay close to a neutral portfolio with $q=0.5N$.

\item The second risk is linked to the price at which the shares are bought or sold:
the trader knows that, at time $T$, his or her portfolio will consist of either $0$ or $N$ shares depending on $S_T$, and the price the trader
pays to buy and sell the shares is random. Being averse to price risk encourages
the trader to have a portfolio that evolves in the same direction
as the price, as is the case in the Bachelier model.
\end{itemize}
Several values of $\gamma$ are considered in Figures \ref{gamma1}
and \ref{gamma2} to see these two effects. Figure \ref{gamma1} shows that the second effect dominates for small values of $\gamma$. When $\gamma$ is really small, the trader is not really interested in hedging and he or she just wants to minimize the cost of delivering the shares (if the option is exercised). Therefore, the hedging strategy is smooth for very small values of $\gamma$. As $\gamma$ increases, the hedging strategy follows the price movement more and more, like the Bachelier $\Delta$: this is the second risk. Now, to see the first effect, we need to consider high values of $\gamma$. Figure \ref{gamma2} shows that when $\gamma$ increases above a certain threshold, then the hedging strategy becomes more and more conservative and ``close'' to $0.5N$: the trader does not want to buy too much because he or she is afraid of being forced to sell afterwards.

\vspace{1.5cm}
In terms of prices, the effect however is unambiguous. Table 6.4 shows that the more risk averse the trader is, the more he or she charges for the risk.

\begin{table}[H]
\centering
\label{tabgamma}
\begin{tabular}{|c|c|c|c|c|c|c|c|}
\hline
$\gamma$  & $1\cdot10^{-8}$ & $2\cdot10^{-8}$ & $5\cdot10^{-8}$ & $2\cdot10^{-7}$ & $1\cdot10^{-6}$ & $2\cdot10^{-6}$ & $5\cdot10^{-6}$\tabularnewline
\hline
Price of the call & 1.955  & 1.968  & 1.994  & 2.060  & 2.207  & 2.308  & 2.521 \tabularnewline
\hline
\end{tabular}
\caption{Prices of the call option for different values of $\gamma$  (tree-based approach).}
\end{table}

\paragraph{Drift and interest rates}

\begin{figure}[H]
\centering \includegraphics[width=0.68\textwidth]{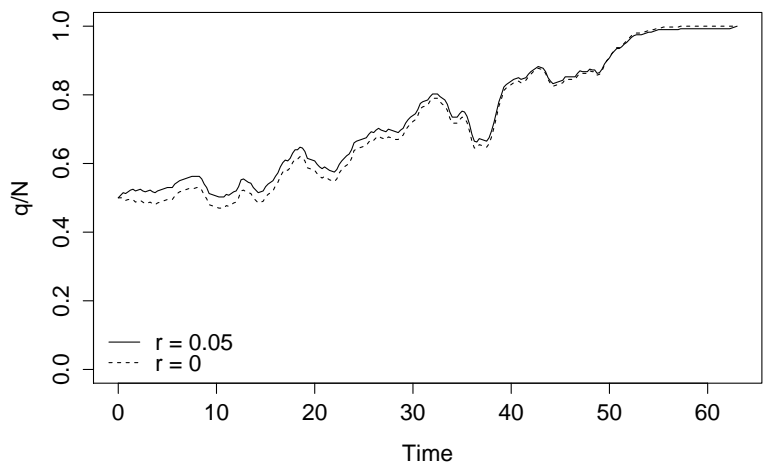} \caption{Optimal portfolio for different values of $r$  (tree-based approach).}
\label{r}
\end{figure}

We have discussed the role of the main parameters. To finish this section on comparative statics, we focus on the respective roles of $r$ and $\mu$. In the classical Bachelier (or Black-Scholes) setting, there is no place for the drift of the underlying asset because the payoff can be replicated. Here, the situation is different. There is indeed, in addition to the hedging problem, another problem of portfolio management in which the trader has to choose the optimal repartition between cash and stock. Figure \ref{r}, where $\mu= 0$, shows that an increase of $r$ from $0\%$ to $5\%$ leads to two effects. As in the classical theory, an increase in the interest rate leads to more shares in the hedging portfolio. This is what we observe, except at the end of the period. The second effect, explaining the behaviour near time $T$, is a pure portfolio management effect. Because a cash position is profitable ($r=5\%$) compared to a long position in the stock, the trader puts less weights on stocks compared to the situation $r=0$.

\begin{figure}[H]
\centering \includegraphics[width=0.68\textwidth]{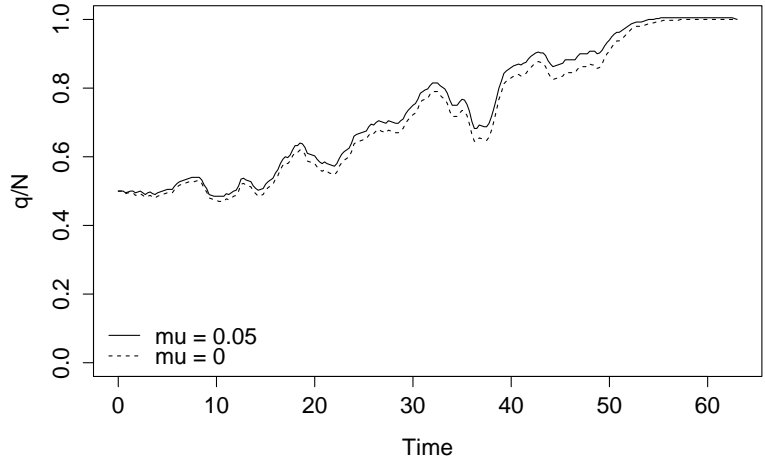} \caption{Optimal portfolio for different values of $\mu$  (tree-based approach).}
\label{mu}
\end{figure}

The same portfolio management effect is at play as far as $\mu$ is concerned. When $\mu$ increases, holding shares is more profitable and the hedging portfolio contains more shares (see Figure \ref{mu}).

\subsection{The difference between physical and cash settlement}

\begin{figure}[H]
\centering \includegraphics[width=0.7\textwidth]{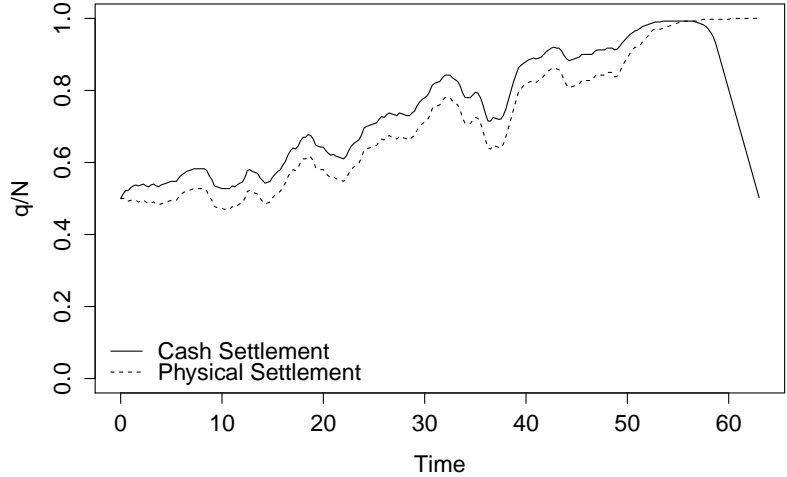} \caption{The difference between a physical settlement and a cash settlement  (tree-based approach).}
\label{cashc}
\end{figure}

One important difference between our model and most of the models in the literature is that we differentiate between physical and cash settlements. To illustrate this point, we use the reference scenario again but with a participation constraint $\rho_m = 50\%$. Figure \ref{cashc} shows that there is an important difference between the two types of settlements when the nominal is large. In both cases, the optimal strategy consists of buying (selling) when the price of the underlying asset is moving up (down) to hedge the position. Hence, when the price $S_t$ is far above $K$ for $t$ close to $T$, the hedging portfolio contains a large number of shares. In the case of a physical delivery, this is fine because the trader has to deliver $N$ shares at expiry if the price stays above $K$. However, in the case of a cash settlement, the trader needs to deliver cash. Figure \ref{cashc} shows that, in order to have cash (and in fact to liquidate the position), the trader progressively sells his or her shares near expiry (given the final cost function we considered, the trader continues to sell with a participation rate to the market equal to $50\%$ after time $T$). To explain what happens far from the expiry date, we rely on a second and more subtle effect. Because the portfolio contains a lot of shares near expiry if $S$ is high, the actual payoff depends on $S$ not only through $(S-K)_+$ but also through $\ell(q)$, which depends implicitly on $S$. Hence, if one applies the classical reasoning, the hedging strategy consists in buying more shares of the underlying stock when the price is increasing. This is the rationale for the difference we observe between the two strategies in Figure \ref{cashc}, except near time $T$.

Table 6.5 shows that a cash settlement is more expensive than a physical settlement. The underlying reason is the final liquidation cost when the option expires in the money.

 \begin{table}[H]
\centering
\begin{tabular}{|c|c|c|c|c|}
\hline
 & Cash settlement $\rho_m = 50\%$  & Physical settlement $\rho_m = 50\%$\tabularnewline
\hline
Price of the call  &  2.401  & 2.100 \tabularnewline
\hline
\end{tabular}
\caption{Prices of the call option for cash delivery and physical delivery  (tree-based approach).}
\end{table}

\subsection{Comparison with the Bachelier model}

Proposition \ref{bounds} argues that the price in the Bachelier
model is lower than the price in our model when $\mu=r=k=0$. This is natural
because our model includes additional costs linked to liquidity. An important
point then is to understand what happens in practice when one uses
the Bachelier model and has to pay the execution costs when rebalancing the $\Delta$-hedging portfolio (at discrete points in time). This scenario highlights the fundamental trade-off between a low
mis-hedge (when $\Delta$-hedging is done at high-frequency) and
low execution costs (when $\Delta$-hedging is done at low-frequency).

The formula for the $\Delta$ in a Bachelier model (when $r=0$) is:
$$
\Delta_{t}^{B}  =  \mathbb{P}\left[\left.S_{T}\geq K\right|S_{t}\right] =  \Phi\left(\dfrac{S_{t}-K}{\sigma\sqrt{T-t}}\right)
$$

In order to carry out a fair comparison between our model and the Bachelier model, we consider several frequencies for the $\Delta$-hedging process.

Let $t_0 = 0,\ldots, t_i =i \delta t, \ldots, t_M= M\delta t$ be a subdivision of $[0,T]$. If at time $t_{i}$ the $\Delta$ of the Bachelier model is $\Delta_{t_{i}}^{B}$, then
we assume that the difference in $\Delta$ (\emph{i.e.},
$\Delta_{t_{i}}^{B}-\Delta_{t_{i-1}}^{B}$) is executed using a perfect
TWAP algorithm over the period $[t_{i},t_{i+1}]$. In other words,
the execution speed is\footnote{We assume that $q_{0}=\Delta_{0}^{B}$.%
}:
\begin{eqnarray*}
v_{t} & = & \begin{cases}
v_{(0)}:=\dfrac{q_{t_{1}}-q_{0}}{\delta t}=\dfrac{\Delta_{0}^{B}-q_{0}}{\delta t}=0 & \text{if }t<t_1\\
v_{(i)}:=\dfrac{q_{t_{i+1}}-q_{t_{i}}}{\delta t}=\dfrac{\Delta_{t_{i}}^{B}-\Delta_{t_{i-1}}^{B}}{\delta t} & \text{for }t\in[t_{i},t_{i+1}),\ 1\le i<M.
\end{cases}
\end{eqnarray*}

Over each time interval $[t_{i},t_{i+1})$ the price obtained by the trader
(excluding the execution costs) is the TWAP over the period:
\begin{eqnarray*}
\mathrm{TWAP}_{i,i+1} & = & \dfrac{1}{\delta t}\int_{t_{i}}^{t_{i+1}}S_{t}dt.
\end{eqnarray*}

A classical result on Brownian bridges leads to the fact that $\mathrm{TWAP}_{i,i+1}|\{S_{t_{i}},S_{t_{i+1}}\}$
is Gaussian with:
\[
\mathbb{E}[\mathrm{TWAP}_{i,i+1}|S_{t_{i}},S_{t_{i+1}}]=\dfrac{S_{t_{i}}+S_{t_{i+1}}}{2}\quad\text{and}\quad\mathbb{V}[\mathrm{TWAP}_{i,i+1}|S_{t_{i}},S_{t_{i+1}}]=\dfrac{\sigma^{2}\delta t}{12}.
\]

Now, the execution costs can be computed as:
$$
\int_{0}^{T}L\left(\frac{v_{t}}{V}\right)Vdt = \sum_{i=0}^{M-1}\int_{t_{i}}^{t_{i+1}}L\left(\frac{v_{t}}{V}\right)Vdt =  V\delta t\sum_{i=1}^{M-1}L\left(\frac{v_{(i)}}{V}\right)
$$
For the terminal condition, we consider the case of a physical settlement, and the terminal condition of Remark 4 with the value of $\rho_m$ of the reference scenario.

\begin{figure}[H]
\centering \includegraphics[width=0.7\textwidth]{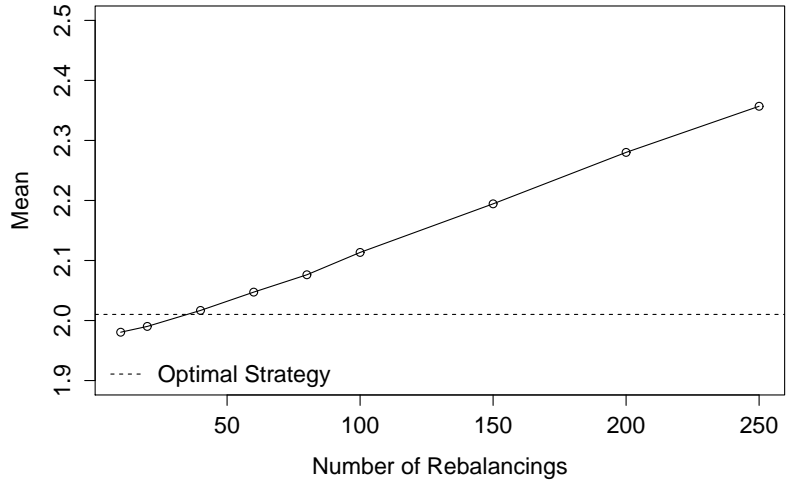} \caption{Average cost of the hedging strategy.}
\label{mean}
\end{figure}

\begin{figure}[H]
\centering \includegraphics[width=0.7\textwidth]{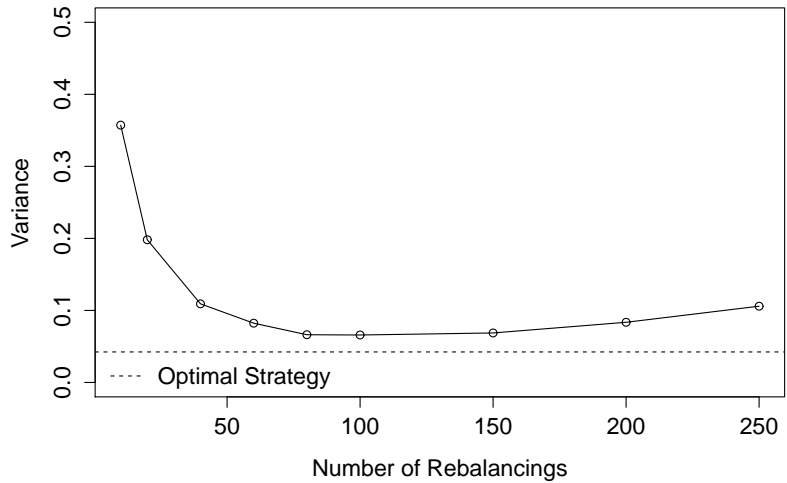} \caption{Variance of the cost of the hedging strategy.}
\label{variance}
\end{figure}

To obtain statistics on the PnL when using the Bachelier model, we consider a Monte-Carlo algorithm with 10,000 draws.
We draw each trajectory $(S_{t_{i}})_{i}$ for the price on the time grid $(t_{i})_{i}$. The trajectories are drawn by using Gaussian increments with a standard deviation parameter $\sigma \sqrt{\delta t}$. Then, we draw values for the TWAPs, and we compute a sample PnL associated with our strategy for the sample trajectory $(S_{t_{i}})_{i}$. The mean and variance of the PnL (in fact $-$PnL) for several values of $M$ (the number of portfolio rebalancings) are given in Figures \ref{mean} and \ref{variance}. To compare the outcomes from the Bachelier model to the outcomes from our model, the same Monte-Carlo procedure is used with 253 points in time (four points per day), but the number of shares to be bought or sold over each period $[t_{i},t_{i+1})$  is computed using the PDE method to approximate $\theta$.

As expected, Figure \ref{mean} shows that the cost of $\Delta$-hedging increases with the frequency of the rebalancings. The level of the average costs when one uses our model corresponds approximately to $M=40$ rebalancings. However, the variance of our strategy is very small compared to the variance of a strategy that consists of $M=40$ rebalancings with the Bachelier $\Delta$. In fact, the variance of the PnL associated with our strategy is smaller than the variance of the PnL associated with the Bachelier strategy for any values of the number of rebalancings. In the case of $\Delta$-hedging, the variance indeed decreases with $M$ for small values of $M$ but reaches a minimum value (greater than the variance of our strategy) and then increases for large values of $M$ because of the presence of execution costs, which generates variance.

\subsection{Numerical examples with permanent market impact}

\begin{figure}[H]
\centering \includegraphics[width=0.7\textwidth]{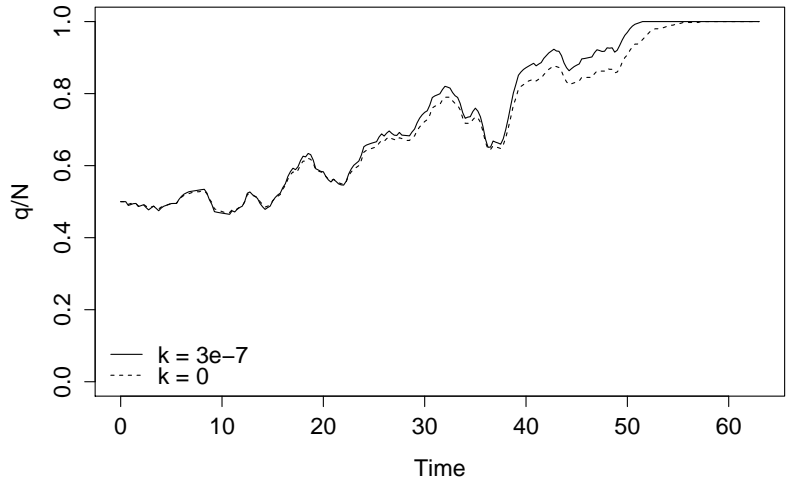} \caption{Optimal strategy with and without permanent market impact.}
\label{pi}
\end{figure}

\begin{figure}[H]
\centering \includegraphics[width=0.7\textwidth]{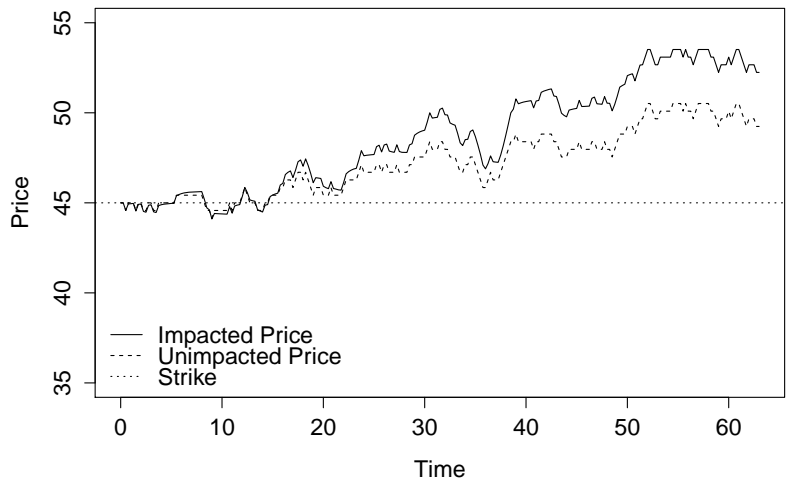} \caption{Impact of the strategy on the price of the underlying asset.}
\label{pi_price}
\end{figure}

So far, we have only considered the case where there is no permanent market impact. As Proposition 4 argues, adding permanent market impact only changes the final condition of the PDE. We use the finite difference scheme to solve the PDE of our reference case when $k=3\cdot 10^{-7}$. The optimal strategies are given in Figure \ref{pi}, and the impacted prices are represented in Figure \ref{pi_price}.

Taking the permanent market impact into account leads to buying more rapidly when the price goes up and selling more rapidly when the price goes down. In fact, there are several effects at play.
\begin{itemize}
\item The first effect is mechanical: when the price of the underlying asset goes up, the position in the underlying asset goes up and it pushes the price of the underlying asset up. Conversely, when the price of the underlying asset goes down, the position in the underlying asset goes down and that pushes the price of the underlying asset down.
\item The second effect is strategic: the trader is risk averse and he or she prefers to know whether he or she will have to liquidate at time $T$. Hence when the trader's position is above a certain threshold, and the price is far above the strike, he or she can buy to push the price up to decrease the level of uncertainty.
    \item Further, because of the permanent market impact, the trader might be tempted to sell shares to push the price down so that the option expires worthless. We do not observe this effect in Figure \ref{pi} (in our experiments, it occurs sometimes when the price nears the strike).
\end{itemize}

In terms of prices, we obtain the results presented in Table 6.6. The table shows that the difference is substantial because the price at which shares are bought is higher and the price at which shares are sold is lower.

 \begin{table}[H]
\centering
\begin{tabular}{|c|c|c|c|c|}
\hline
 & $k=0$& $k=3\cdot10^{-7}$\tabularnewline
\hline
Price of the call  &  2.067  & 2.689 \tabularnewline
\hline
\end{tabular}
\caption{Prices of the call option with and without permanent market impact (PDE approach).}
\end{table}

\section*{Conclusion}

In this paper, we present a new model to price and hedge options
in the case of an illiquid underlying asset or when the nominal of the option is too large
to neglect the execution costs. We show that the price of a call option
when the execution costs are taken into account is the solution of a 3-variable
nonlinear PDE that can be solved using various numerical techniques.
The comparisons to the use of classical models show the relevance
of our approach. Although our paper focuses on the case of a call
option, it can be generalized to other options, with or without
physical delivery at maturity. For instance, \cite{gueantpuroyer}
uses a similar framework to price an Accelerated Share Repurchase
contract -- an execution contract that can be seen as a Bermudan option with an Asian payoff.

\vspace{3em}

\section*{Appendix: Proofs}

Proof of Lemma \ref{xqS}:\\

By definition, $$d(e^{-rt}X_t) = e^{-rt}\left(-v_tS_t dt - V_t L\left(\dfrac{v_{t}}{V_{t}}\right) dt\right).$$

Therefore,
$$ e^{-rT}X_T - e^{-rt}x =  \int_t^T e^{-rs}\left(-v_sS_s ds - V_s L\left(\dfrac{v_{s}}{V_{s}}\right) ds\right)$$
$$= e^{-rt}qS -  e^{-rT}q_T S_T + \int_t^T q_s e^{-rs}\left(-rS_s ds + \mu ds + \sigma dW_s \right)-  \int_t^T  e^{-rs}V_s L\left(\dfrac{v_{s}}{V_{s}}\right) ds,$$ where we used an integration by parts.

Reorganising the terms results in:

$$X_{T}+q_{T} S_{T} = e^{r(T-t)}(x+qS)$$$$ +  e^{r(T-t)} \left(\int_t^T e^{-r(s-t)} q_s(\mu-rS_s)ds + \int_t^T e^{-r(s-t)} q_s \sigma dW_s - \int_t^T e^{-r(s-t)} V_{s}L\left(\dfrac{v_{s}}{V_{s}}\right) ds  \right).$$\qed\\

Proof of Proposition \ref{bounds}:\\

We define
\begin{eqnarray*}
I(t,q,S,v) & = & \int_t^T e^{-r(s-t)} q_s(\mu-rS_s)ds + \int_t^T e^{-r(s-t)} q_s \sigma dW_s\\
                    & & - \int_t^T e^{-r(s-t)} V_{s}L\left(\dfrac{v_{s}}{V_{s}}\right) ds - e^{-r(T-t)}\Pi(q_T,S_T),\\
\end{eqnarray*}
and
\begin{eqnarray*}
w(t,q,S) & = & \inf_{v\in\mathcal{A}_{t}}J_{t}(q,S,v) = \inf_{v\in\mathcal{A}_{t}} \mathbb{E}\left[\exp\left(-\gamma e^{r(T-t)}I(t,q,S,v)\right)\right].
\end{eqnarray*}

It is straightforward to verify that $I(t,q,S,0)$ has a Laplace transform defined on $\mathbb{R}$. Therefore,

$$w(t,q,S)  \le \mathbb{E}\left[\exp\left(-\gamma e^{r(T-t)}I(t,q,S,0)\right)\right] < + \infty.$$

This proves that $\theta$ is bounded from above.

Coming to the other inequality, Jensen's inequality gives:
\begin{eqnarray*}
w(t,q,S) & \ge & \inf_{v\in\mathcal{A}_{t}}\exp\left(-\gamma e^{r(T-t)}\mathbb{E}[I(t,q,S,v)]\right)\\
 & \ge & \exp\left(-\gamma e^{r(T-t)}\mathbb{E}\left[   \int_t^T e^{-r(s-t)} q_s(\mu-rS_s)ds  - e^{-r(T-t)}N(S_T-K)_+ \right]\right)
\end{eqnarray*}
Because $q$ is bounded and because $S_s$ is a Gaussian random variable for all $s$, $w(t,q,s)>-\infty$. Therefore, $\theta$ is also bounded from below.\\

If $\mu=r=0$, then
$$\theta(t,q,S) = \frac{1}{\gamma}\log(w(t,q,S)) \ge  N\mathbb{E}\left[(S_{T}-K)_+\right].$$\qed\\

Proof of Proposition \ref{convex}:\\

We first prove that $I$ is a concave function of $(v,q)$. Given $t\in\left[0,T\right]$ and $S\in\mathbb{R}$, the functions

\begin{eqnarray*}
(q,v) & \mapsto & \int_{t}^{T}e^{-r(s-t)}q_{s}^{t,q,v}(\mu - r S_s) ds=\int_{t}^{T}e^{-r(s-t)}\left(q+\int_{t}^{s}v_{u}du\right)(\mu - r S_s) ds\\
\end{eqnarray*}

and

\begin{eqnarray*}
(q,v) & \mapsto & \int_{t}^{T}e^{-r(s-t)}q_{s}^{t,q,v}\sigma dW_{s}=\int_{t}^{T}e^{-r(s-t)}\left(q+\int_{t}^{s}v_{u}du\right)\sigma dW_{s}\\
\end{eqnarray*}

are linear and hence concave.

Because $L$ is convex, the following function is concave:

\begin{eqnarray*}
(q,v) & \mapsto & -\int_{t}^{T}e^{-r(s-t)} V_{s}L\left(\dfrac{v_{s}}{V_{s}}\right)ds.\\
\end{eqnarray*}

Because $\Pi(q_T,S_T)$ is a convex function of $q_T = q + \int_t^T v_s ds$,

\begin{eqnarray*}
(q,v) & \mapsto & I(t,q,S,v)\\
\end{eqnarray*}

is a concave function (sum of four concave functions).\\

To go on, we need a lemma (which is a consequence of H\"older inequality):
\begin{lem}
\label{logexp} Let $X$ and $Y$ be two random variables. Let $\lambda\in\left[0,1\right]$. The following inequality holds:
\begin{eqnarray*}
\log\mathbb{E}\left[\exp\left(\lambda X+\left(1-\lambda\right)Y\right)\right] & \leq & \lambda\log\mathbb{E}\left[\exp\left(X\right)\right]+\left(1-\lambda\right)\log\mathbb{E}\left[\exp\left(Y\right)\right]
\end{eqnarray*}
\end{lem}

Let us now recall that
\begin{eqnarray*}
\theta(t,q,S) & = & \dfrac{e^{-r(T-t)}}{\gamma}\inf_{v\in\mathcal{A}_{t}}\log\mathbb{E}\left[\exp\left(-\gamma e^{r(T-t)}I(t,q,S,v)\right)\right].
\end{eqnarray*}

If $t\in\left[0,T\right]$, $S\in\mathbb{R}$, $\hat{q},\check{q}\in\mathbb{R}$
and $\lambda\in\left[0,1\right]$, then for $\hat{v},\check{v}\in\mathcal{A}_{t}$,
we have the following inequality:
\begin{eqnarray*}
\theta(t,\lambda\hat{q}+\left(1-\lambda\right)\check{q},S) & \leq & \dfrac{e^{-r(T-t)}}{\gamma}\log\mathbb{E}\left[\exp\left(-\gamma e^{r(T-t)}I\left(t,\lambda\hat{q}+\left(1-\lambda\right)\check{q},\lambda\hat{v}+\left(1-\lambda\right)\check{v},S\right)\right)\right]
\end{eqnarray*}
Using Lemma \ref{logexp} and the concavity of $I$, we have:
\begin{eqnarray*}
\theta(t,\lambda\hat{q}+\left(1-\lambda\right)\check{q},S) & \leq & \lambda\dfrac{ e^{-r(T-t)}}{\gamma}\log\mathbb{E}\left[\exp\left(-\gamma e^{r(T-t)}I\left(t,\hat{q},\hat{v},S\right)\right)\right]\\
 &  & +\left(1-\lambda\right)\dfrac{e^{-r(T-t)}}{\gamma}\log\mathbb{E}\left[\exp\left(-\gamma e^{r(T-t)}I\left(t,\check{q},\check{v},S\right)\right)\right].
\end{eqnarray*}
Because this inequality holds for all $\hat{v},\check{v}\in\mathcal{A}_{t}$,
we can take the infima over them on the right-hand side:
\begin{eqnarray*}
\theta(t,\lambda\hat{q}+\left(1-\lambda\right)\check{q},S) & \leq & \lambda\dfrac{e^{-r(T-t)}}{\gamma}\inf_{\hat{v}\in\mathcal{A}_{t}}\log\mathbb{E}\left[\exp\left(-\gamma e^{r(T-t)}I\left(t,\hat{q},\hat{v},S\right)\right)\right]\\
 &  & +\left(1-\lambda\right)\dfrac{e^{-r(T-t)}}{\gamma}\inf_{\check{v}\in\mathcal{A}_{t}}\log\mathbb{E}\left[\exp\left(-\gamma e^{r(T-t)}I\left(t,\check{q},\check{v},S\right)\right)\right]\\
 & \le & \lambda\theta(t,\hat{q},S)+\left(1-\lambda\right)\theta(t,\check{q},S),
\end{eqnarray*}
which proves the Proposition.\qed\\

Proof of Proposition \ref{withoutperm}:\\

Given the hypotheses, it is classical to prove that $u$ is a viscosity solution of:
\begin{eqnarray*}
-\partial_{t}u - \mu \partial_S u -\dfrac{1}{2}\sigma^{2}\partial_{SS}^{2}u-\sup_{|v|\le \rho_m V_t}\left\{ v\partial_{q}u+\left(rX-vS-L\left(\dfrac{v}{V_{t}}\right)V_{t}\right)\partial_{x}u\right\}  & = & 0,
\end{eqnarray*}

Let us consider $\varphi\in C^{1,1,2}((0,T)\times\mathbb{R}\times\mathbb{R})$
and $(t^{*},q^{*},S^{*})$ such that:
\begin{itemize}
\item $\theta^{*}-\varphi$ has a local maximum at $(t^{*},q^{*},S^{*})$,
\item $\theta^{*}(t^{*},q^{*},S^{*})=\varphi(t^{*},q^{*},S^{*})$.
\end{itemize}
Let us define $\psi(t,x,q,S)=-\exp\left[-\gamma e^{r(T-t)}\left(x+qS-\varphi(t,q,S)\right)\right]\in C^{1,1,1,2}((0,T)\times\mathbb{R}\times\mathbb{R}\times\mathbb{R})$.

Because $\theta(t,q,S)=x+qS+\frac{e^{-r(T-t)}}{\gamma}\log(-u(t,x,q,S))$, $\forall x^{*}\in\mathbb{R}$, $(t^{*},x^{*},q^{*},S^{*})$ is such
that $u_{*}-\psi$ has a local minimum at $(t^{*},x^{*},q^{*},S^{*})$.

Using the super-solution property of $u$, we obtain:
\[
\begin{array}{lcr}
\partial_{t}\psi(t^{*},x^{*},q^{*},S^{*})+\mu \partial_{S}\psi(t^{*},x^{*},q^{*},S^{*}) +\dfrac{1}{2}\sigma^{2}\partial_{SS}^{2}\psi(t^{*},x^{*},q^{*},S^{*})\\
+\sup_{|v|\le \rho_m V_t}\left\{ v\partial_{q}\psi(t^{*},x^{*},q^{*},S^{*})+\left(rx^*-vS^*-L\left(\dfrac{v}{V_{t}}\right)V_{t}\right)\partial_{x}\psi(t^{*},x^{*},q^{*},S^{*})\right\}  & \le & 0.
\end{array}
\]
As $\psi(t,x,q,S)=-\exp\left[-\gamma e^{r(T-t)}\left(x+qS-\varphi(t,q,S)\right)\right]$,
we have:
\begin{itemize}
\item $\partial_{t}\psi(t,x,q,S)=\gamma e^{r(T-t)}\psi(t,x,q,S)\partial_{t}\varphi(t,q,S) + \gamma r e^{r(T-t)}\psi(t,x,q,S) \left(x+qS-\varphi(t,q,S)\right) $
\item $\partial_{x}\psi(t,x,q,S)=-\gamma e^{r(T-t)} \psi(t,x,q,S)$
\item $\partial_{q}\psi(t,x,q,S)=-\gamma e^{r(T-t)}\psi(t,x,q,S)(S-\partial_{q}\varphi(t,q,S))$
\item $\partial_{S}\psi(t,x,q,S)=-\gamma e^{r(T-t)} \psi(t,x,q,S)(q-\partial_{S}\varphi(t,q,S))$
\item $\partial_{SS}^{2}\psi(t,x,q,S)=\gamma^{2} e^{2r(T-t)}\psi(t,x,q,S)(q-\partial_{S}\varphi(t,q,S))^{2}+\gamma e^{r(T-t)}\psi(t,x,q,S)\partial_{SS}^{2}\varphi(t,q,S))$
\end{itemize}
Hence:
\begin{eqnarray*}
0 & \ge & -\gamma e^{r(T-t^*)}\psi(t^{*},x^{*},q^{*},S^{*})\Bigg(-\partial_{t}\varphi(t^{*},q^{*},S^{*}) -r(q^*S^*-\varphi(t^{*},q^{*},S^{*}))\\
&& + \mu(q^*-\partial_S\varphi (t^{*},q^{*},S^{*})) -\dfrac{1}{2}\sigma^{2}\partial_{SS}^{2}\varphi(t^{*},q^{*},S^{*})-\dfrac{1}{2}\gamma\sigma^{2}e^{r(T-t^*)}(q^*-\partial_{S}\varphi(t^{*},q^{*},S^{*}))^{2}\\
 &  & +\inf_{|v|\le \rho_m V_t}\left\{-v\partial_{q}\varphi(t^{*},q^{*},S^{*})-L\left(\dfrac{v}{V_{t}}\right)V_{t}\right\} \Bigg).\\
\end{eqnarray*}
Therefore:
\begin{eqnarray*}
0 & \ge & -\partial_{t}\varphi(t^{*},q^{*},S^{*}) +r \varphi(t^{*},q^{*},S^{*}) + q^*(\mu-rS^*)  -\mu \partial_S\varphi (t^{*},q^{*},S^{*})\\
&& -\dfrac{1}{2}\sigma^{2}\partial_{SS}^{2}\varphi(t^{*},q^{*},S^{*})-\dfrac{1}{2}\gamma\sigma^{2}e^{r(T-t^*)}(q^*-\partial_{S}\varphi(t^{*},q^{*},S^{*}))^{2}+V_t H(\partial_{q}\varphi(t^{*},q^{*},S^{*})).\\
\end{eqnarray*}
This proves that $\theta$ is a sub-solution of the equation.

The same reasoning applies to the super-solution property and this
proves the result.\qed\\

\bibliographystyle{plain}

\begin{thebibliography}{10}
\bibitem{almgreninit} R. Almgren, N. Chriss, Optimal execution of
portfolio transactions. Journal of Risk, 3, 5-40, 2001.

\bibitem{almgrenciti}R. Almgren, C. Thum, E. Hauptmann, H. Li. Direct estimation of equity market impact. Risk, 18(7):58–62, 2005.


\bibitem{sbank}


P. Bank, D. Baum, Hedging and portfolio optimization in financial
markets with a large trader, Mathematical Finance, 14, 1-18, 2004.

\bibitem{bardi} M. Bardi, I. Capuzzo-Dolcetta, Optimal control and
viscosity solutions of Hamilton-Jacobi-Bellman equations, Springer,
2008.

\bibitem{bouchard} B. Bouchard, G. Loeper, Almost-sure hedging with permanent price impact, working paper, 2015.


\bibitem{tcbarlessoner} G. Barles, H. M. Soner, Option pricing with
transaction costs and a nonlinear Black-Scholes equation, Finance
and Stochastics, 2, 369-397, 1998.

\bibitem{scetin}


U. Çetin, R. Jarrow, P. Protter, Liquidity risk and arbitrage pricing
theory, Finance and Stochastics, 8, 311-341, 2004.

\bibitem{scetin2}


U. Çetin, R. Jarrow, P. Protter, M. Warachka, Pricing options in an
extended Black-Scholes economy with illiquidity: theory and empirical
evidence, The Review of Financial Studies, 19, 493-529, 2006.

\bibitem{cetin3}


U. Çetin, L.C. Rogers, Modelling liquidity effects in discrete time,
Mathematical Finance, 17, 15-29, 2007.

\bibitem{cetingamma} U. Çetin, H. M. Soner, N. Touzi, Option hedging
for small investors under liquidity costs, Finance and Stochastics,
14:317-341, Jan 2010.

\bibitem{tccvitanic1} J. Cvitanić, I. Karatzas, Hedging and portfolio
optimization under transaction costs: a martingale approach, Mathematical
Finance, 6, 133-165, 1996.

\bibitem{tccvitanic2} J. Cvitanić, H. Pham, N. Touzi, A closed-form
solution to the problem of super-replication under transaction costs,
Finance and Stochastics, 3(1), 35-54, 1999.

\bibitem{ferretti} M. Falcone, R. Ferretti. Semi-Lagrangian schemes for Hamilton–Jacobi equations, discrete representation formulae and Godunov methods. \emph{Journal of computational physics}, 175(2), 559-575, 2002

\bibitem{gatheral}  J. Gatheral, No-Dynamic-Arbitrage and Market Impact, Quantitative Finance, Vol. 10, No. 7, pp. 749-759, 2010


\bibitem{gueantpuroyer} O. Guéant, J. Pu, G. Royer, Pricing and hedging
of Accelerated Share Repurchase, to appear in IJTAF, 2015

\bibitem{gueant}
O. Guéant, Optimal execution and block trade pricing: a general framework,
to appear in Applied Mathematical Finance, 2015

\bibitem{gueantperm}

O. Guéant, Permanent market impact can be nonlinear, working paper,
2013.

\bibitem{jaimungal} S. Jaimungal, D. Kinzebulatov, D. Rubisov, Optimal
Accelerated Share Repurchase, working paper, 2013.

\bibitem{lehalle1} C.-A. Lehalle, S. Laruelle, R. Burgot, S. Pelin,
M. Lasnier, Market Microstructure in Practice. World Scientific publishing,
2013.

\bibitem{lehalle2} C.-A. Lehalle, M. Lasnier, P. Bessson, H. Harti,
W. Huang, N. Joseph, L. Massoulard, What does the saw-tooth pattern
on US markets on 19 july 2012 tell us about the price formation process.
Crédit Agricole Cheuvreux Quant Note, Aug. 2012.

\bibitem{tcleland} H. E. Leland, Option pricing and replication with
transactions costs. The Journal of Finance, 40(5), 1283-1301, 1985

\bibitem{lialmgren} T. M. Li, R. Almgren, A Fully-Dynamic Closed-Form
Solution for $\Delta$-Hedging with Market Impact, to appear in Operations
Research, 2013.

\bibitem{longstaff} F. A. Longstaff, Optimal portfolio choice and
the valuation of illiquid securities, The Review of Financial Studies,
14, 407-431, 2001.

\bibitem{mono}

M.G. Crandall, P.-L. Lions, Two approximations of solutions of Hamilton-Jacobi equations, \emph{Math. Comp.} 43, 1-19, 1984


\bibitem{fplaten}


E. Platen, M. Schweizer, On feedback effects from hedging derivatives,
Mathematical Finance, 8, 67-84, 1998.

\bibitem{rogerssingh} L. C. Rogers, S. Singh, The cost of illiquidity
and its effects on hedging. Mathematical Finance, 20(4), 597-615,
2010.

\bibitem{st}

W. Schachermayer, and J. Teichmann. How close are the option pricing formulas of Bachelier and Black–Merton–Scholes? Mathematical Finance, 18(1), 155-170, 2008.


\bibitem{schied}


A. Schied, T. Schöneborn, M. Tehranchi. Optimal basket liquidation
for CARA investors is deterministic. Applied Mathematical Finance,
17(6), 471-489, 2010.

\bibitem{fschon}


P. J. Schönbucher, P. Wilmott, The feedback effects of hedging in
illiquid markets, SIAM Journal on Applied Mathematics, 61, 232-272,
2000.

\bibitem{fsircar}


R. Sircar, G. Papanicolaou, Generalized Black-Scholes models accounting
for increased market volatility from hedging strategies, Applied Mathematical
Finance, 5(1), 45-82, 1998.\end{thebibliography}

\end{document}